# Using 4D STEM to probe mesoscale order in molecular glass films prepared by physical vapor deposition

*Debaditya Chatterjee[1,‡], Shuoyuan Huang[1], Kaichen Gu[2,†], Jianzhu Ju[2], Junguang Yu[3], Harald Bock[4], Lian Yu[3], Mark D. Ediger[2], and Paul M. Voyles[1,*]*

[1]Department of Materials Science and Engineering, University of Wisconsin-Madison, Madison, WI 53706, USA

[2]Department of Chemistry, University of Wisconsin-Madison, Madison, WI 53706, USA

[3]School of Pharmacy, University of Wisconsin-Madison, Madison, WI 53705, USA

[4]Centre de Recherche Paul Pascal - CNRS & Université de Bordeaux, 33600 Pessac, France

KEYWORDS: molecular glasses, scanning transmission electron microscopy, physical vapor deposition, nanoscale ordering

ABSTRACT: Physical vapor deposition can be used to prepare highly stable organic glass systems where the molecules show orientational and translational ordering at the nanoscale. We have used low-dose four-dimensional scanning transmission electron microscopy (4D STEM), enabled by a fast direct electron detector, to map columnar order in glassy samples of a discotic mesogen using a 2 nm probe. Both vapor deposited and liquid cooled glassy films show domains of similar orientation, but their size varies from tens to hundreds of nanometers, depending on processing. Domain sizes are consistent with surface diffusion mediated ordering during film deposition.

These results demonstrate the ability of low-dose 4D STEM to characterize mesoscale structure in a molecular glass system which may be relevant to organic electronics.

Physical vapor deposition (PVD) can synthesize stable glassy phases of many materials,[1–4] but for organic molecules, PVD can prepare a more unusual material: glasses in which the molecules show overall orientational ordering.[5] This anisotropy in structure causes anisotropy in optical,[6] mechanical,[7] and electronic properties,[8] making the films potentially useful for organic electronics.[9] Discotic mesogens are attractive for applications in organic electronic devices as they can show columnar ordering via π–π stacking of molecules, one-dimensional charge mobility,[10] and alignment relative to the substrate that can be tailored for various applications.[8] Phenanthro[1,2,3,4-ghi]perylene-1,6,7,12,13,16-hexacarboxylic 6,7,12,13-tetraethyl, 1,16-dimethyl ester ("phenanthroperylene ester", henceforth) is a discotic mesogen designed for applications in organic heterojunction devices. It forms a glassy phase due to the presence of a distorted arene core which suppresses crystallization while maintaining electronic properties, and it has a relatively low molecular weight, which enables PVD at moderate evaporation temperatures.[11] It forms a hexagonal columnar liquid crystal phase upon cooling from the isotropic liquid phase at 519 K and undergoes a glass transition at 393 K.[12]

Previous research on molecular glass films has shown that ordering during PVD is mediated by film surface and the surface-vacuum interface during growth[13] and that varying PVD conditions can create various orientation-ordered structures. Increasing the substrate temperature is equivalent to decreasing the deposition rate, a phenomenon called the rate-temperature superposition (RTS),[14–16] and the RTS shift factor depends on the surface mobility. One orientation-ordered structure of phenanthroperylene ester glass involves the discotic molecules orienting edge-on to the surface and organizing into columns via π–π stacking, as illustrated in Figure 1(a). The columns then organize into hexagonal arrays. The x-ray correlation length for the π–π stacking (q ~ 1.8 Å$^{-1}$) ranges from 1.5 nm to 3.3 nm, (~4 – 9 molecules), and the correlation length for the hexagonal packing of the columns (q ~ 0.4 Å$^{-1}$) ranges from 5.1 nm to 33 nm, which is ~3 – 20 molecular diameters, but the in plane structure is isotropic over macroscopic length scales. X-ray scattering is insensitive to local mesoscale ordering in these samples, especially in plane. 4D STEM nanodiffraction is sensitive to nano- and mesoscale ordering in glasses,[17–19] and recent developments in fast electron cameras made possible low-dose experiments on beam-sensitive organic materials.[20–22] However, low dose has limited the spatial resolution to 5 to 10 nm and 4D STEM on organic materials has focused on very pronounced nanocrystalline ordering, which produces bright, sharp diffraction features that are easy to detect at low dose.

Here, we use an ultrafast, high sensitivity direct electron detector[23] to study the mesoscale orientational order in thin films of a phenanthroperylene ester from the scattering signal from the stacking of molecules via π-bonding. We achieve results uninfluenced by beam damage at ~3 nm resolution, which is sufficient to image orientation domains as small as 10 nm, even for the subtle, diffuse scattering features from a glass. The resulting orientation domain maps are consistent with surface diffusion control of the domain size and quantify structural features like domain misorientation that are important for functional properties.

Figure 1(a) shows the 4D STEM experimental schematic. The diffraction arcs at k ~ 0.286 Å$^{-1}$, visible in Figure 1(c) and figure S1(a), arise from π–π stacking between the discotic molecules with a spacing of ~3.5 Å, which is consistent with previous x-ray scattering investigations[12,16,24] on this material showing a strong peak at q ~1.8 Å$^{-1}$ (q = 2πk). The orientation of the arcs therefore reflects the in-plane orientation of the columns of discs. Figure 1(b) shows a spatial map of the orientation, expressed as an angle between 0° and 180° with respect to horizontal, for a $T_s$ = 392 K sample, derived by processing many diffraction patterns as described in the supporting information. Each angle is assigned a color from the cyclic color scale, and the arrows overlaid on the map also show the local orientation. Domains 10s of nanometers across have similar orientations and extend through the 40 nm thickness of sample. The nanodiffraction data are a 2D projection of the structure through the thickness in the beam direction. Most of the nanodiffraction patterns look like the top and bottom patterns in Figure 1(c), with only one set of arcs indicating only one domain through the thickness. The middle pattern in Figure 1(c) comes from the boundary Additional structural features like local disorder and out-of-plane columnar orientation can be extracted from the 4D STEM datasets as discussed in figure S2.

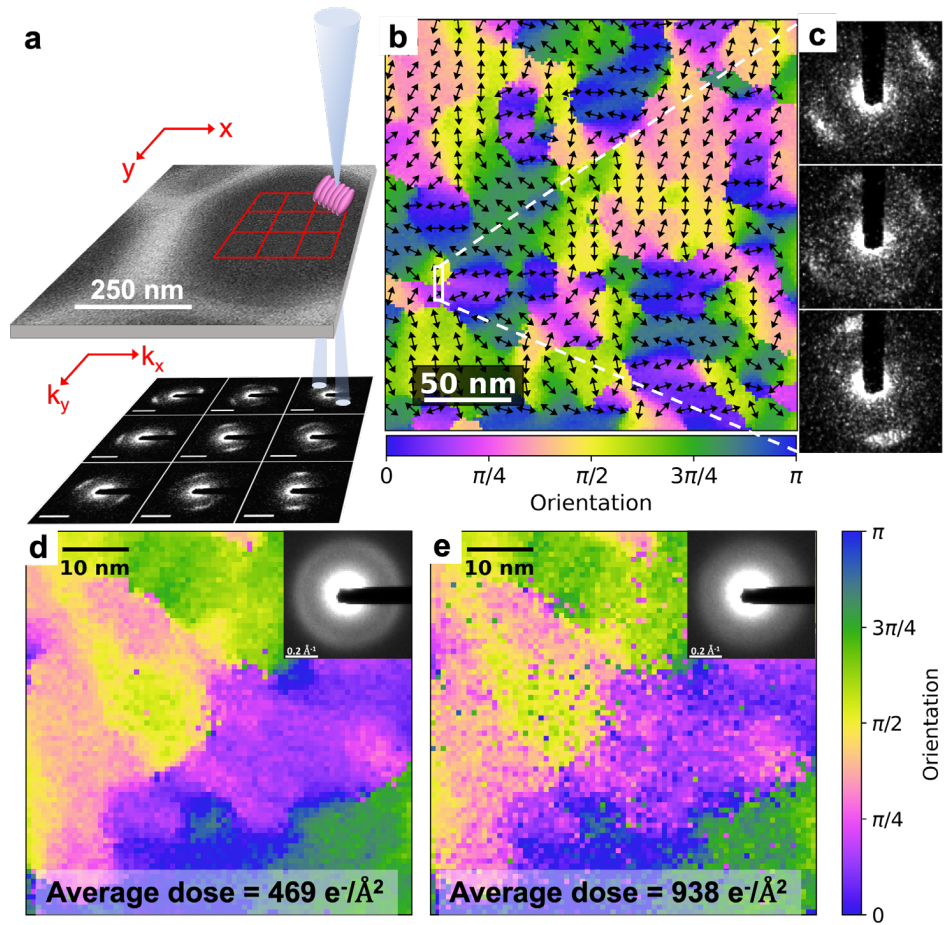

**Figure 1:** (a) 4D STEM experiment schematic – a diffraction pattern is acquired from each position as shown with the grid of patterns below the sample. The pink column of discs below the probe is a schematic of the columnar stacking and the diffraction geometry. (b) An orientation map from a 250 nm region on a $T_s$ = 392 K sample constructed from the angle of the diffraction arcs at each position on the sample. Each angle is color-coded from the cyclic color scale shown, which reveals domains having similar orientation within the sample. The arrows overlaid on the map also show the local orientation. (c) Diffraction patterns (spatially averaged and Gaussian filtered to reduce noise from the region in the white rectangle on the map highlighting variation in the diffraction signal between neighboring domains (top and bottom) and the boundary (middle)) (d) and (e) show orientation maps at accumulated dose of 469 e$^-$/Å$^2$ and 938 e$^-$/Å$^2$. Insets: Computed averages of the 4D STEM nanodiffraction patterns acquired with accumulated dose of 469 e$^-$/Å$^2$ and 938 e$^-$/Å$^2$ respectively. The main datasets in this work were acquired at ~117 e$^-$/Å$^2$ or lower.

The phenanthroperylene ester molecules are damaged by exposure to the electron beam.[25,26] The orientation maps in figures 1(d) and (e) acquired at doses of 469 e$^-$/Å$^2$ and 938 e$^-$/Å$^2$ show domains with similar sizes and shapes, but the map acquired after a higher dose is noisier because damage has reduced the visibility of the diffraction arc features and thus the reliability of the data analysis. The rest of the orientation maps presented here were acquired at ~117 e$^-$/Å$^2$ or lower, so they have minimal impact from beam damage. Further analysis of beam damage on these samples is presented in figure S3.

Figure 2 shows example orientation maps from samples prepared at different processing conditions. Figure 2(a), (b) and (c) from vapor deposited samples and reveal orientation domains that roughly triple in size as the substrate temperature is increased from 370 K to 392 K. Figure 2(d) and 2(e) show orientation maps from a film that was deposited, then heated to close to the liquid-crystal ordering temperature, then cooled back into a glass. This liquid-cooled glass sample showing much larger domains extending over several hundred nanometers. They also show typical liquid crystal defects. Points in the map around which the orientation varies smoothly are focal conics in a planar alignment, a defect in columnar hexagonal liquid crystals and smectic A liquid crystals.[27,28] There have been previous polarized microscopy,[29] AFM,[30] and TEM[31] reports on the structure and length scale of focal conic domains, but this is the first observation at nanometer-scale spatial resolution. Additional orientation maps are shown in supplementary figure S4(a)-S7(a). 4D STEM data from samples deposited at $T_s$ = 335 K and $T_s$ = 355 K (not shown) exhibit a continuous ring in nanodiffraction, not arcs, so they either do not have strong in-plane ordering, or the in-plane ordering domain size is small compared to the 40 nm sample thickness.

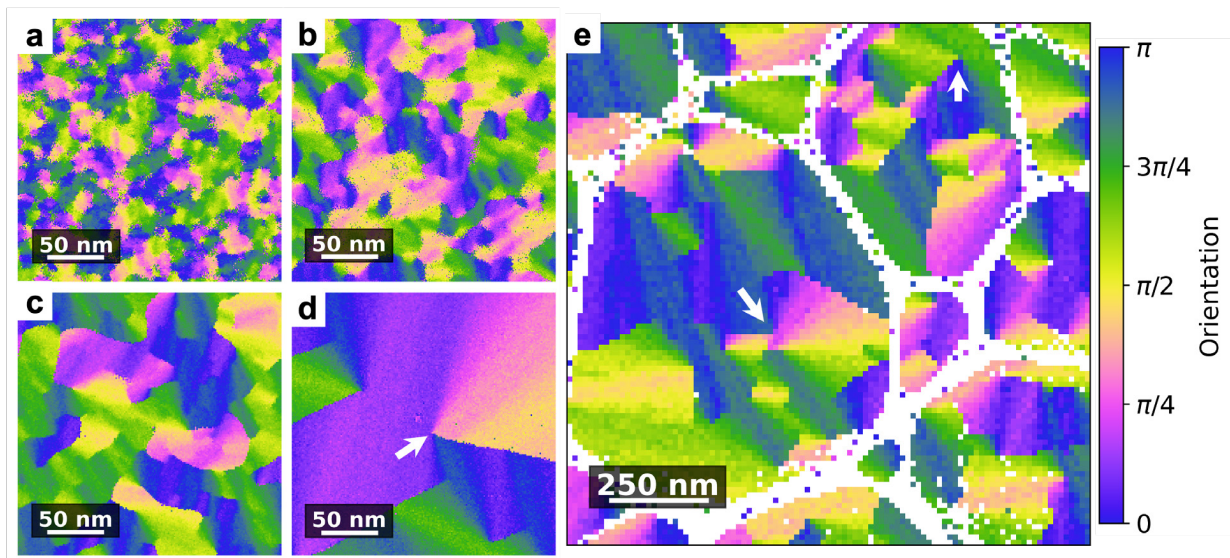

**Figure 2:** Orientation maps for samples deposited at (a) $T_S$ = 370 K (0.94 $T_g$), (b) $T_S$ = 380 K (0.97 $T_g$), (c) $T_S$ = 392 K ($T_g$), all with a 252 nm field of view. (d) and (e) orientation maps for a liquid-cooled glass with 252 nm and 1.38 μm fields of view respectively. Arrows point out focal conics. The white regions in 2(e) are substrate carbon laces.

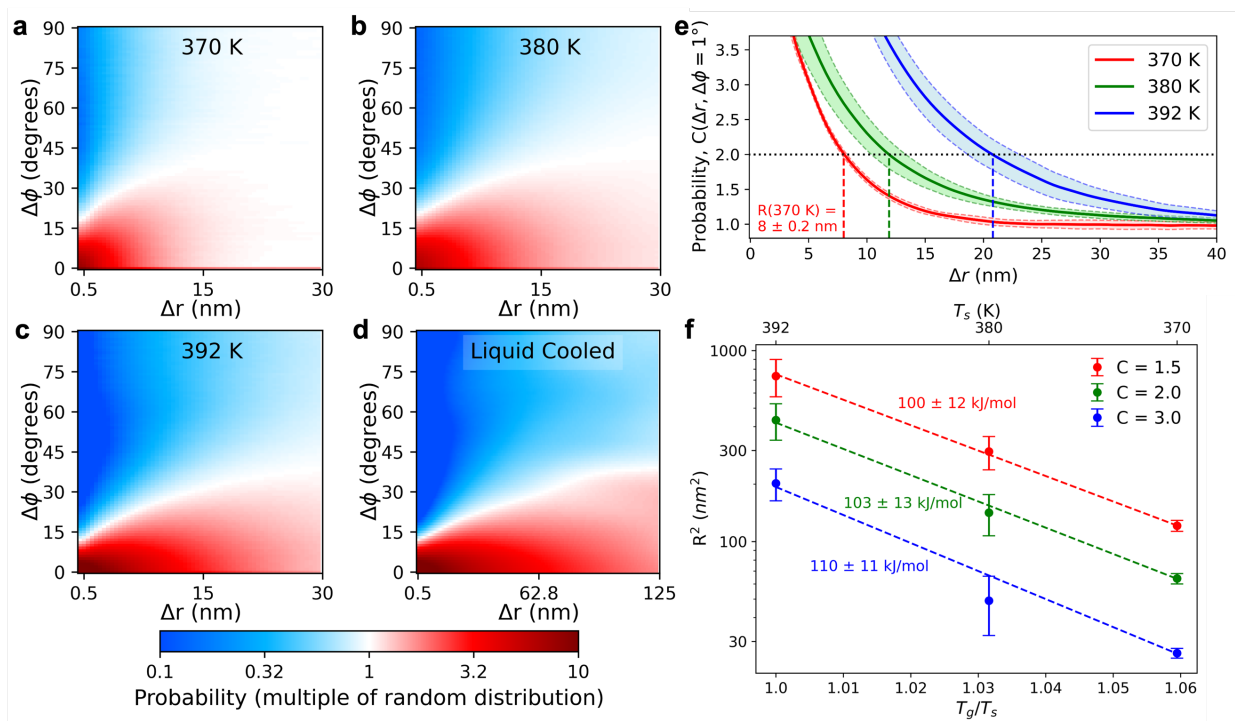

**Figure 3:** Orientation correlation C as a function of distance ($\Delta r$) and misorientation ($\Delta \phi$) for the (a) $T_s$ = 370 K, (b) $T_s$ = 380 K, (c) $T_s$ = 392 K, and (d) liquid cooled samples. The length scale of orientation in each sample can be visualized through the extent of the red region (C>1). In the PVD samples, the orientations become uncorrelated at $\Delta r$ < 50 nm, but for the liquid cooled sample the orientation persists even at $\Delta r$ > 100 nm. (e) Slices through C at $\Delta \phi = 1°$ as a function of $\Delta r$, used to determine ordering length, R, for PVD samples. (f) $R^2$ as a function of $T_g/T_s$ showing Arrhenius behavior with activation energies near 105 kJ/mol.

Figure 3 shows the orientation correlation probability, C, of similar columnar orientations at a distance ($\Delta r$) and misorientation ($\Delta \phi$), averaged over the several orientation maps, for the $T_s$ = 370 K, 380 K, 392 K, and liquid cooled samples. $C$ is given by[20]

$$C(\Delta r, \Delta \phi) = \frac{\langle I(r,\phi)I(r+\Delta r,\phi+\Delta\phi)\rangle_{r,\phi}}{\langle I(r,\phi)\rangle^2_{r,\phi}},$$

where $\langle \cdots \rangle_{r,\phi}$ indicates averaging over probe positions and azimuthal angle in nanodiffraction. It is normalized to the probability for a random distribution of columnar orientations, so the large value near the origin indicates strong ordering within the orientation domains. We extract an ordering length scale R, related to a domain size, by C($\Delta r$) at fixed $\Delta \phi$=1°, the minimum misorientation calculated, shown in figure 3(e). The solid traces are mean C values from multiple datasets for each sample, and the shaded regions are uncertainties estimated as the standard deviation of the mean (see figures S4(b)-S7(b)). R is defined as the $\Delta r$ value at fixed value of C(>1) for each sample. Figure 3(e) illustrates this process for C($\Delta r$, $\Delta \phi$=1°) = 2, which results in R = 8.0 ± 0.2 nm, 11.9 ± 1.4 nm, and 20.8 ± 2.2 nm for the 370 K, 380 K, and 392 K samples, respectively. Uncertainties in R are estimated from the uncertainty in C. Figure 3(f) shows that the ordering lengths follow an Arrhenius relationship with activation energy near 105 kJ/mol, independent of the choice of C used to determine R.

The orientation domain size of the as-deposited films is controlled by surface diffusion of the molecular orientation during growth. Surface diffusivity for many glasses is orders of magnitude faster than bulk diffusivity at temperatures near $T_g$, like the substrate temperatures used here.[32] High surface diffusivity means that newly-deposited molecules can sample many configurations during the surface residence time, $t_{res}$, before they are buried by more deposition. Sampling many configurations allows energetically favorable states to be found.[1] Once the molecules are buried, they can only access slow bulk dynamics, so the surface configuration of each layer is frozen in. Thus, annealing a bulk sample does not show similar ordering to depositing a film at the same temperature. The film surface is also important because the film-vacuum interface creates a preferred molecular orientation,[13] which plays a larger role in determining the final molecular orientations and the domain sizes, than the substrate-film interface.[5]

The physical picture of anisotropic film growth dominated by surface diffusion is as follows: A layer of molecules is deposited. They reorient fast enough that most of the molecules are aligned with the discs edge-on to the substrate (rather than parallel) by the energy of the film surface-vacuum interface. The discs coalesce to form columns, but the in-plane orientation of the columns is not controlled by the surface-vacuum interface. Domains of similar in-plane column orientation form, and the domain boundaries can shift via changes in the orientation of the discs at the boundary. The surface layer is slowly buried by more deposition, freezing its molecules in place because bulk diffusivity is low. As a result, the domain size is controlled by the surface diffusivity that controls molecular orientation. The domains extend through the thickness of the film because the columns in the surface layer align themselves with the columns in the layers below them to form domains. Even in samples where the extent of in-plane orientation creates domains smaller than 40 nm, the orientation propagates through the thickness of the film due to a templating effect by the lower deposited layers. However, at low enough temperature, the domain size is small enough that columnar growth is not sustained through the entire film thickness.

This picture is consistent with both the order of magnitude of the domain size and the temperature dependence of the domain size. We can estimate the absolute domain size from the surface diffusion length for molecular orientation during deposition of a monolayer as $R \approx 2\sqrt{D_s(T_s)t_{res}}$. $R$ is the domain size, and $D_s$ is the surface diffusivity. Based on previous work by Bishop et. al.[14,16] measuring the birefringence as a function of deposition rate at 392 K, we estimate the relaxation time for molecular orientation at the surface at 392 K to be $\tau \sim 2.4$ ms. The surface diffusivity from the Debye-Stokes-Einstein equation,[33,34] $D_s = 2r^2/9\tau$, is 50 nm$^2$/s for a molecular radius, $r$, of 0.75 nm for phenanthroperylene. At a deposition rate of 0.025 nm/s, $t_{res}$ is ~ 60 s, the time required to deposit a 1.5 nm thick monolayer (one molecular diameter). These quantities predict $R \approx 100$ nm at 392 K, which is in reasonable agreement with the data in figure 3(e) ($R$ = 27 nm at 392 K, using C=1.5 for example). The activation energy for $D_s$ is 110 kJ/mol, measured from orientational order in x-ray scattering experiments on PVD samples.[16] $R^2 \propto D_s$ in this model, so figure 3(f) shows $R^2$ vs. $1/T_s$, leading to activation energies of 100-110 kJ/mol for various $C$, all in in excellent agreement with the activation energy for $D_s$. The agreement of both absolute diffusivity and temperature dependence is strong evidence that surface diffusion during PVD is responsible for the growth of ordered columnar domains.

The ordering length in the 4D STEM orientation maps (10s to 100s of nm) is much larger than the scattering correlation length (2-5 nm) determined from the inverse peak width in either TEM diffraction (figure S8) or x-ray scattering (previous studies), showing the added value of the direct determination of the mesoscale structure by 4D STEM. The correlation length depends on fluctuations in both the π–π separation between discs and their orientation. The orientation map domain size depends on the orientation of the π–π separations, but is less sensitive to variable in the distance. The larger scale persistence of domain orientation is likely to be important for charge transport,[8] so access to this mesoscale information is critical.

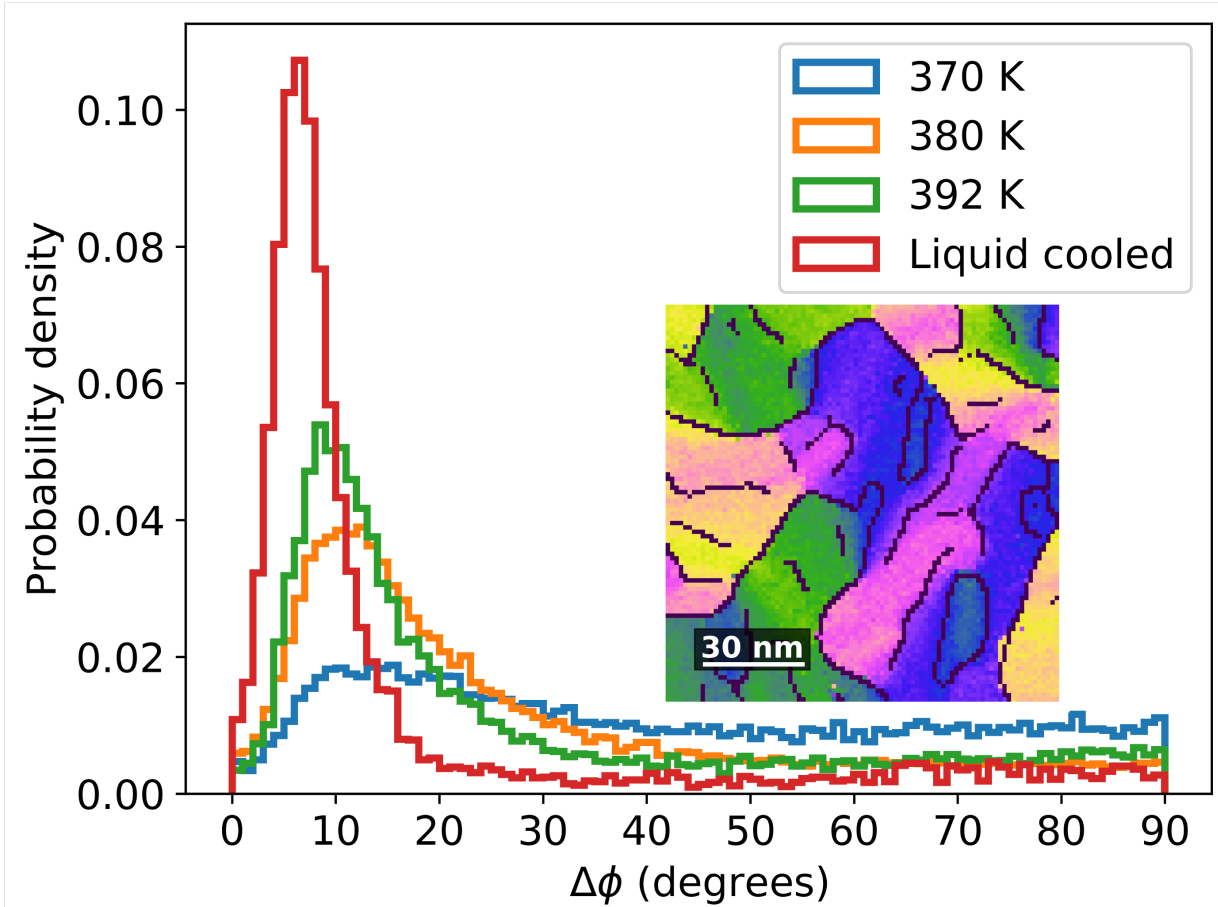

**Figure 4:** Boundary misorientation distributions for T$_S$ = 370K, 380 K, 392 K, and liquid-cooled glass samples. (Inset) Orientation map from a 392 K sample showing domain boundaries determined using the Canny edge detection algorithm (see supporting information).

Misorientation across domain boundaries also plays a role in transport.[8] Figure 4 shows the distribution of domain misorientations, defined as a difference in local orientation across a domain boundary, determined from the orientation maps. Low misorientation angles are favored for all samples. The distributions shift towards lower misorientations and become narrower for the higher substrate temperature film, and the liquid-cooled glass sample. The peaks are at 15º, 11°, 9°, and 7° for $T_s$ = 370 K , $T_s$ = 380 K, $T_s$ = 392 K, and liquid-cooled glass samples respectively, and the half-widths at half-maximum are 17°, 9°, 6°, and 3° for the same samples. Experiments at a lower dose (60 e⁻/Å²) show distributions with similar widths, so the contribution of beam damage to figure 4 is small. The preference for low misorientation boundaries and the curved nature of the domain boundaries in PVD samples suggest that low misorientation boundaries have lower in interfacial energy than higher misorientation boundaries. In general, films with larger domains and lower misorientation boundaries should have higher charge mobility, so PVD may provide access to higher mobility films than can be achieved via bulk processing at the same temperature.

These results demonstrate the value of 4D STEM with a high-speed camera like the Direct Electron Celeritas, which enables a high degree of control over electron dose and spatial sampling.[35] The Celeritas runs at up to ~86,000 fps and has high gain and low noise for single electron sensitivity.[23] With a 0.6 pA probe, an exposure time of 8 ms (125 fps) is sufficient to record enough signal, depositing a dose of ~100 e⁻/Å², which allows a small probe step (~2 nm) for improved spatial resolution and lets us map larger regions on the sample without significant sample drift during the acquisition time. Other experiments[20] have used probe steps larger than the probe diameter to minimize dose and beam damage, which we can also do for large domains (see Figure 2(e)).

Finally, are these materials really glasses? They exhibit very strong structural order in several ways, with a preferred overall orientation for the molecules in one direction and local preferred orientations in another. The nanodiffraction could come from a textured polycrystalline material, as could the orientation domain maps. The liquid-cooled sample even has well-defined line defects! It is helpful to make a distinction between glasses and amorphous materials. Calling a material a glass is a statement about its kinetics: Does it exhibit a glass transition on heating and cooling? Calling a material amorphous is a statement about structure: Does it lack long-range order, especially in wide-area diffraction? These phenanthroperylene ester films do exhibit a glass transition in calorimetry,[12] so they are glasses despite their structural order. Questions like "How structurally ordered can a material be and still be a glass?" and "How do these materials change their structural order through the glass transition?" are interesting topics for future research.

In summary, low-dose, high speed 4D STEM experiments show that vapor-deposited glassy thin films of a discotic liquid crystal molecule form domains with a local preferred in-plane orientation for the π–π stacking direction of the discs. The average domain size increases with substrate temperature, and the average misorientation across domain boundaries decreases. The domain size is controlled by the surface diffusivity for reorientation of the discs: The film/vacuum interface promotes an edge-on orientation for the discs, which stack into columns, and the in-plane orientation of the columns is determined by diffusion of the domain boundaries by reorientation of discs. Large in-plane domains and low domain misorientations are likely to favor high charge mobility, so the capability to control the structural order of such materials by direct deposition, without high temperature annealing, may be beneficial for organic electronics. This study sets the stage for further investigations into fundamental ordering processes in organic systems.

**Supporting Information**

Supporting Information is available in a separate document describes additional experimental details and results covering electron dose and diffraction correlation lengths.

**Acknowledgements**

Primary funding for the work was provided by the Wisconsin MRSEC (NSF DMR-1720415). The authors gratefully acknowledge Carter Francis (University of Wisconsin – Madison) for developing the Seq-IO library, contributing to HyperSpy and PyXem, and for helpful discussions. The authors also thank Colin Ophus (National Center for Electron Microscopy, Lawrence Berkeley National Laboratory, Berkeley, CA, USA) for helpful discussions and help with orientation correlation analysis using py4DSTEM.

**Corresponding Author**

Paul Voyles – Department of Materials Science and Engineering, University of Wisconsin-Madison, Madison, WI 53706, USA

Email: paul.voyles@wisc.edu

**Present Address**

‡Applied Materials Inc., Santa Clara, California, USA

†Exponent, New Territories, Hong Kong SAR, China

1. **Author Contributions**

PMV and ME conceived the study. DC did the 4D STEM experiments and analyzed the data. SH and DC wrote the analysis software. KG and JJ grew the films. JY, DC, and LY processed the films to obtain the liquid cooled sample. HB synthesized the phenanthroperylene molecules. ME, LY, and DC developed the diffusion growth model. DC and PMV wrote the manuscript. All authors helped develop the results and conclusions and edited the manuscript.

**Data Availability**

Table of contents entry:

High-resolution, low-dose 4D STEM measurements show that physical vapor deposited (PVD) films of a glassy organic semiconductor contain nanoscale domains with varying in-plane orientation. The domain size is controlled by surface diffusion, so it can be controlled with PVD growth conditions.

Debaditya Chatterjee, Shuoyuan Huang, Kaichen Gu, Jianzhu Ju, Junguang Yu, Harald Bock, Lian Yu, M.D. Ediger, Paul M. Voyles[*]

**Using 4D STEM to probe mesoscale order in molecular glass films prepared by physical vapor deposition**

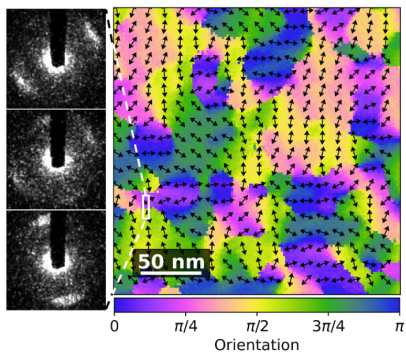

# Supporting Information

## Using 4D STEM to probe mesoscale order in molecular glass films prepared by physical vapor deposition

*Debaditya Chatterjee, Shuoyuan Huang, Kaichen Gu, Jianzhu Ju, Junguang Yu, Harald Bock, Lian Yu, Mark D. Ediger, and Paul M. Voyles[*]*

*Email: paul.voyles@wisc.edu

## 1. Methods

### 1.1 Material synthesis

The phenanthroperylene ester was synthesized as described in Kelber *et al.*[1] The melting point of its crystalline solid is $T_m$ = 520 K. The molecules are discotic and the supercooled liquid exhibits a hexagonal columnar phase below 492 K that undergoes a glass transition at $T_g$ = 392 K.[2]

### 1.2 TEM sample preparation via physical vapor deposition

Thin films of phenanthroperylene ester were prepared in a custom-built deposition chamber as outlined in Gujral *et al.*[2] using the same method as detailed in Bishop *et al.*[3] The films were deposited directly onto ultrathin carbon/lacey carbon TEM grids (Ted Pella, product number 01824: ultrathin carbon film on lacey carbon support film, 400 mesh, copper) or on lace-free ultrathin carbon films (Electron Microscopy Sciences, CF400-Cu-UL: 3-4 nm ultrathin carbon film on 400 mesh, copper). These grids were chosen due to their high electron transmittance and amorphous structure resulting in an isotropic featureless background which does not interfere with the sample diffraction signal (see figure S8(b) for substrate diffraction profile). Depositions were performed at a rate of 0.25 Å/s at substrate temperatures of 335 K (0.85 $T_g$) 355 K (0.9 $T_g$), 370 K (0.94 $T_g$), 380 K (0.97 $T_g$), and 392 K ($T_g$). This parameter space was chosen as the substrate temperature has a much larger impact on structure than the available range of deposition rates: decreasing the deposition rate by an order of magnitude is equivalent to increasing the substrate temperature by 17 K in terms of the orientational ordering by the RTS principle[3]. The scattering geometry is also an important consideration for electron diffraction experiments, and we chose from the deposition rate and substrate temperature range where previous two-dimensional GIWAXS investigations had shown a high degree of in-plane order via intermolecular π–π stacking[3] which is the only structure we can probe with the 4D STEM experimental design used here. The spacing between the discotic molecules is ~3.5 Å which causes a diffraction peak at k ~ 0.286 Å$^{-1}$ (q ~ 1.8 Å$^{-1}$). The thickness of the deposited films was chosen to be 40 nm as the STEM electron transmittance was measured to be 73% at this thickness which is ideal for nanodiffraction experiments.[4] A liquid-cooled glass sample was prepared from a thin film sample vapor deposited at 380 K. Phenanthroperylene ester forms a liquid crystal phase near 450 K,[5] so we heat the vapor deposited film to 447 K from room temperature at 150 K/min, and then up to 457 K at 1 K/min to form an equilibrium hexagonal columnar liquid crystal phase. After holding at 457 K for 1 min, it was cooled at 2.5 K/min back to room temperature to form the liquid-cooled glass. Temperature was controlled with a Linkam THMS 600 hot stage purged with dry $N_2$. Samples were loaded onto a TEM holder and kept in vacuum overnight to minimize surface contamination prior to insertion into the TEM.

**1.3 4D STEM data acquisition**

The 4D STEM orientation mapping experiment schematic is shown in figure 1 and the analysis method is illustrated in figure S1. The sample is scanned using a beam of electrons across a two-dimensional grid of probe positions as shown in red in figure 1(a). A two-dimensional diffraction pattern is recorded at each position, generating a four-dimensional dataset (with two real space (x,

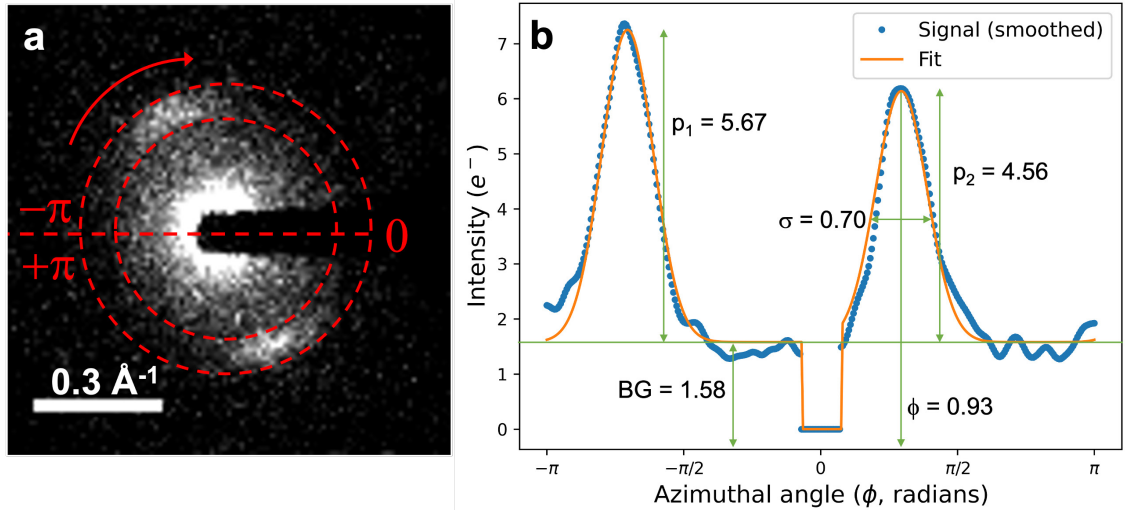

**Supplementary Figure 1:** (a) Representative diffraction pattern from a 4D STEM dataset showing a pair of diffraction peaks on either side of the transmitted beam partially covered by a beam stop. (b) The azimuthal unwrapped intensity profile from the diffraction signal in the selected k-range, along with a fitted profile showing the fitted parameters (constant background (BG), peak 1 intensity ($p_1$), peak 2 intensity ($p_2$), peak position ($\phi$) in the azimuthal range [$-\pi/2,+\pi/2$], and peak width ($\sigma$)) from the analysis.

y) dimensions and two reciprocal space ($k_x$, $k_y$) dimensions). Figure S1(a) shows a typical diffraction pattern after spatial binning. The orientation of the Bragg peak pair reveals the sample's local columnar orientation as illustrated with the pink molecular schematic in figure 1(a), which varies spatially as shown in the grid of diffraction patterns also in figure 1(a).

Datasets for orientation mapping were acquired using a 200 kV Thermo Fisher FEI Titan microscope operating in µP EFSTEM mode with a convergence semi-angle of $\alpha=0.7$ mrad, a C2 aperture sized 10 µm, and a camera length of 1700 mm. The Gaussian-shaped probe size was 2 nm at full-width at half-maximum. During alignment, the sample height, condenser astigmatism and focus were corrected by viewing the Ronchigram using a 70 µm C2 aperture. A spot number of 11 was used yielding a probe current of 0.6 pA. Data was collected using a Direct Electron Celeritas detector[6] with a 256×256 readout area with a pixel size of 0.005 Å$^{-1}$ providing a sensor area of 1.28 Å$^{-1}$×1.28 Å$^{-1}$. A beam stop was used to cover the zero beam in the diffraction pattern to prevent damage to the detector. All data were dark and gain corrected, and intensities lower than

7 in each frame was set to 0 since the detector false positive rate per pixel per frame is >10 ppm for intensities lower than 7, following which intensities were divided by 266 - the average gain of the detector, to convert intensities to electron units.

Since phenanthroperylene ester is a beam sensitive material, locations for data collection were determined via HAADF-STEM scanning with a dwell time of 3 µs at low magnification, following which the beam was blanked, the goniometer was moved by a few micrometers to reach an area not already exposed to the electron beam, and the magnification was increased so that the scan area decreased to the desired region of interest (for example, a STEM magnification of 320 kX set the scan area to 252 nm × 252 nm). Using a probe step of ~0.2 nm, diffraction patterns were acquired from each scan position with a dwell time of 125 µs with the detector running at 8000 fps, immediately after un-blanking the beam. Due to the very high acquisition speed, we do not obtain sufficient signal in a single frame, but it offers the flexibility to spatially average diffraction patterns to improve the signal level at the expense of spatial resolution. For example, a spatial binning factor of 10 would mean summing 10×10 frames, increasing the signal-to-noise ratio by 100 while increasing the effective spatial sampling step size to 2 nm. Post-acquisition analysis has revealed that a spatial bin factor of 8 is optimal for orientation mapping analysis, which has been primarily used for the data presented in this paper. For a 252 nm × 252 nm region of interest, datasets were acquired with 1200×1200 probe positions which was binned down to a spatial size of 150×150. Time required for such an acquisition is 3 min and typical stage drift in this time is 0.9 nm (0.36% of scanned region). Each diffraction pattern in the data was binned by a factor of 2 to obtain 128×128 frames with a pixel size of 0.01 Å$^{-1}$ to reduce the total dataset size and improve signal-to-noise ratio. Acquisition from a large region on the liquid-cooled sample was performed on a 1.38 µm × 1.38 µm region (see figure 2(e)) with a 13.8 nm probe step and exposure time of

8 ms at each step with the detector running at 125 fps, without any post-acquisition spatial binning for this dataset.

The electron dose received by the sample during a single exposure of a 2 nm diameter 0.6 pA probe with an exposure time of 125 μs is 1.49 e$^-$/Å$^2$ and the average dose received during a 4D STEM scan with a probe step of 0.2 nm is 117 e$^-$/Å$^2$ due to overlapping probe positions. This acquisition scheme followed by spatial binning by a factor of 8, is equivalent to using a 1.6 nm probe step and 8 ms exposure time in terms of spatial resolution and signal-to-noise ratio, but since the extent of probe overlap is lower while using a 1.6 nm step size, the dose received by the sample goes down to 95 e$^-$/Å$^2$, which is a better experimental design in terms of damage minimization if the expected signal levels are already known and flexibility between spatial resolution and signal levels is not necessary.

To determine the angular position of the Bragg pair, the intensity is radially integrated within the range k = 0.27-0.33 Å$^{-1}$ as shown by the red dashed rings in figure S1(a) to calculate the intensity as a function of azimuthal angle ($\phi$) as shown in the blue dotted profile in figure S1(b). The azimuthal angle is numbered following a convention marked in (b) ranging from $-\pi$ to $+\pi$ clockwise starting from the center left. The unwrapped intensity profile is fitted to a composite function as shown by the orange trace, consisting of two Gaussians constrained to have equal widths and be $\pi$ radians apart, allowed by the inversion symmetry in the diffracting molecular columns. The region blocked by the beam stop is treated using a mask in the composite fitting function. Since a pair of peaks are fit, only the peak position within a $\phi$ range of $-\pi/2$ to $+\pi/2$ is recorded. The beam stop blocks ~22° of the profile which does not pose a problem with the analysis since we constrain the fitted peaks to be $\pi$ radians apart. The fit parameters are recorded as a function of probe position on the sample for further analysis. The recorded orientations are offset by adding $\pi/2$ to obtain angles in the range 0 to $\pi$ radians. The reported peak widths ($\sigma$) are the FWHM of the fitted Gaussian at each pixel obtained by multiplying the standard deviation fit parameter with a factor of $2\sqrt{2\ln 2}$. The orientation maps are further analyzed to determine misorientations across domain boundaries and sizes of domains. Analysis of microscopy data is performed using Hyperspy (v1.7.1)[7] and Pyxem (v0.14.2)[8] implemented in Python (v3.10), and boundary misorientations are determined using Canny algorithm implemented using Scikit-Image (v0.18.1).[9] Orientation autocorrelation analysis is performed on orientation mapping datasets by the method of Panova et. al.[10] using py4DSTEM[11] (v0.13.6) implemented in Python (v3.8). The decay in orientation correlation as a function of distance at low misorientation is used to estimate ordering lengths for different samples.

### 1.4 TEM diffraction data acquisition

Selected area diffraction (SAD) patterns for beam damage measurements and correlation length calculations were acquired with a parallel beam in TEM mode at 200 kV accelerating voltage using a C2 aperture sized 100 μm on an area on the sample with a diameter of 1.5 μm. The probe current was 68 pA giving a dose rate of 2.405 e⁻/Å²s. The camera length used was 300 mm and data was collected using a Direct Electron Celeritas detector[6] with a 512×512 readout area with a pixel size of 0.0017 Å⁻¹ providing a sensor area of 0.868 Å⁻¹. A beam stop was used to cover the zero beam in the diffraction pattern to prevent damage to the detector. Data was acquired for 60 s at 100 fps. All data were dark and gain corrected, and intensities lower than 7 in each frame was set to 0, following which intensities were divided by 266 which is the average gain of the detector to convert intensities to electron units. These datasets were temporally binned by a factor of 10 to have an exposure time of 0.1 s in each frame.

Locations for data collection are determined in a similar way like we did for 4D STEM data acquisition. A location on the sample showing strong diffraction signal is determined and the beam is blanked. Then the goniometer is shifted by ~10 µm and data acquisition is launched immediately after un-blanking the beam. The diffraction pattern at each time step was azimuthally integrated to obtain intensity profiles as a function of radial k magnitude. Artifacts due to the presence of the beam stop are avoided by setting the pixels under the beam stop to NaN prior to azimuthal integration. These profiles in the k range 0.1-0.4 Å$^{-1}$ were fitted to a sum of a power law ($Ak^r + C$) background and two Gaussians for the substrate carbon diffraction peak and sample peak from π–π stacking diffraction (see figure S8(a) and S8(b)). This analyzed dataset is used to plot the signal-to-background ratio for the π–π stacking peak, normalized by the value at the first time-step, as a function of accumulated dose using the dose rate to convert from the time-step to dose. The fitted peak width for the π–π stacking peak for the profile at the first time-step with minimal beam damage, is used to calculate the correlation length for columnar π–π stacking by taking the inverse of the full-width at half-maximum.

2. **Additional 4D STEM structural maps**

Small orientation disorder broadens the diffraction arcs in the azimuthal direction, analogous to mosaic spread in x-ray diffraction. Figure S2(a) shows a spatial map of the width of the diffraction arcs (σ, in figure S1(b)). Pixels with high width correspond to regions of high disorder in the sample, like domain boundaries which can be seen by comparing figure S2(a) and 1(b). Figure S2(b) shows a spatial map of the ratio of peak intensities ($p_1/p_2$, from figure S1(b)) from the fitted Gaussians at each position. Deviation from 1 could arise from out-of-plane orientation in the diffracting columns.[12] A comparison between figure S2(b) and 1(b) shows that regions within a domain in figure 1(b) show a similar peak ratio value in figure S2(b) indicating a spatial correlation between in-plane and out-of-plane columnar orientations.

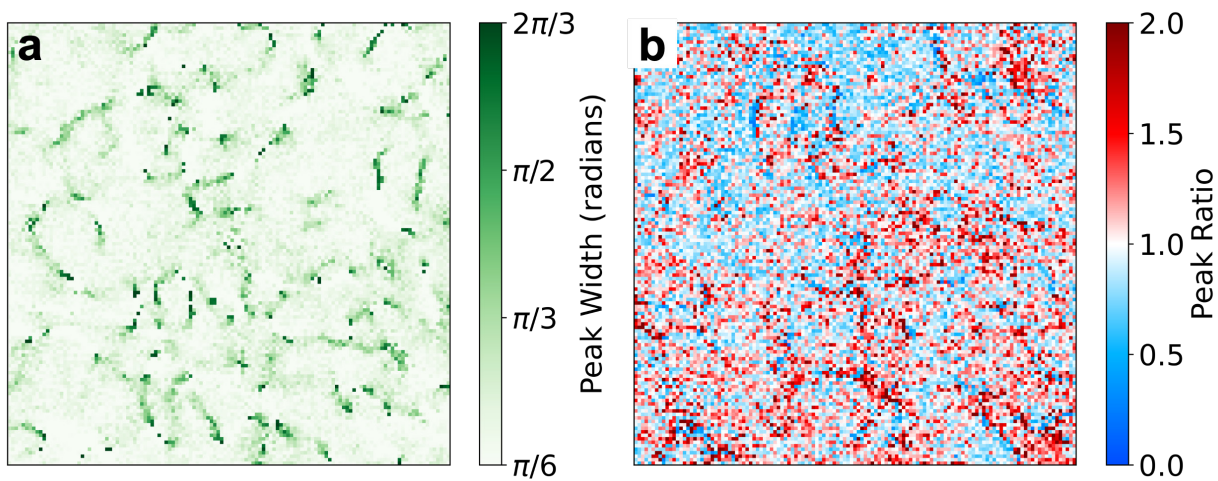

**Supplementary Figure 2.** (a) A spatial map of the width of the diffraction peak in the azimuthal direction at each position. (b) A spatial map of the ratio of peak intensities from the fitted Gaussians at each position.

The rich 4D STEM dataset provides access to a variety of other structural properties of the films which are expected to connect to their functional properties. For example, all of the samples exhibit significant orientation disorder within the domains, reflected in the azimuthal width of the nanodiffraction π–π arcs. The average width for the $T_s$ = 392 K sample is ~40°, as shown in figure S2(a), and the other samples are similar (see figure S3(d)). Since this disorder is observed at a single probe position, it occurs through the thickness of the film. The orientation disorder increases around domain boundaries, as seen by comparing figure 1(b) and S2(a). Nanodiffraction also probes the tilt of the columns of molecules out of plane: a column of molecules exactly parallel to the substrate and therefore exactly edge-on to the beam will diffract equally into both arcs in the pair. (This geometry is analogous to a zone axis crystal, with equal deviation parameter for the two diffraction spots, which creates equal intensities under kinematic conditions.[13] The low scattering power and disorder of this sample should produce nearly kinematic diffraction.) Figure S2(b) shows that the ratio of arc intensities is mostly 1, reflecting the in-plane orientation, but some domains are tilted slightly out of plane, and there are large tilts at the domain boundaries.

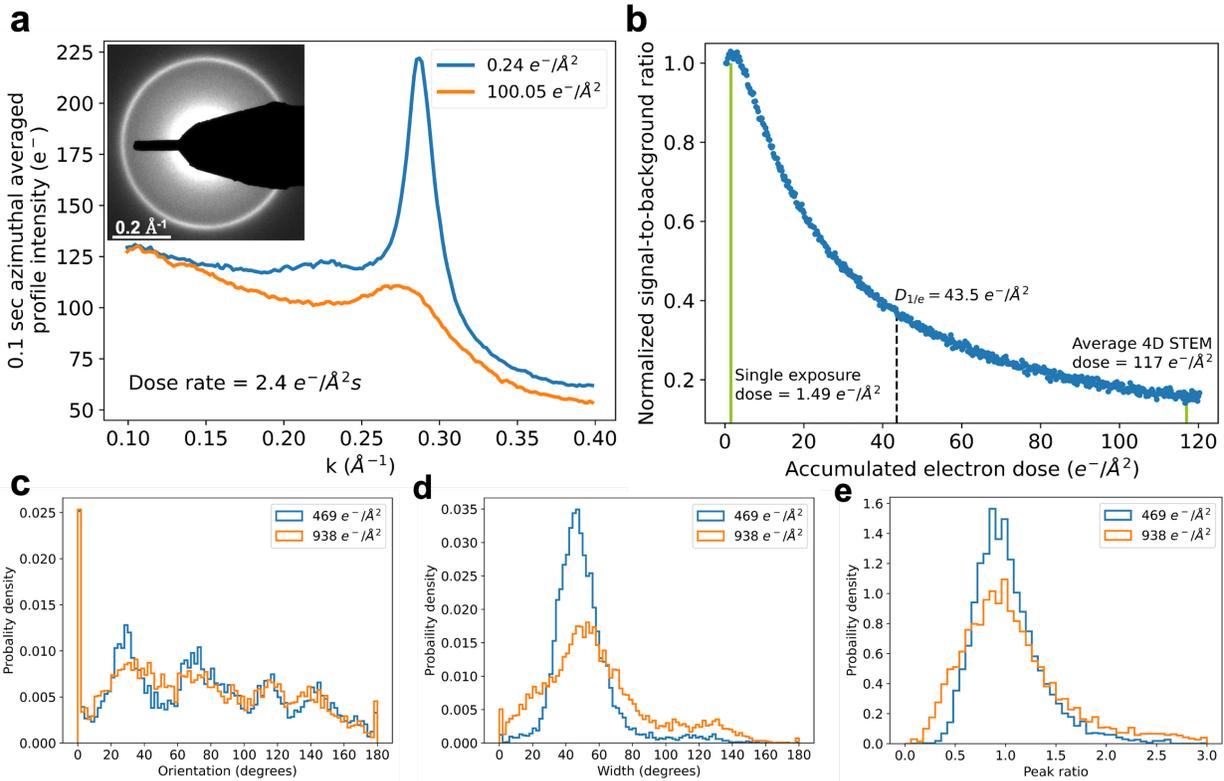

**Supplementary Figure 3:** (a) TEM diffraction profiles as a function of radial k magnitude, averaged in the azimuthal direction. Inset: a TEM diffraction pattern acquired at low accumulated dose showing the ring arising from p-p stacking between molecules at k = 0.286 Å$^{-1}$. (b) Normalized signal-to-background ratio as a function of accumulated electron dose extracted from profiles like the ones in (a). The characteristic dose for this system is determined to be $D_{1/e}$ = 43.5 e$^-$/Å$^2$. Distributions of the (c) orientations, (d) peak widths, and (e) peak intensity ratios at two accumulated dose levels of 469 e$^-$/Å$^2$ and 938 e$^-$/Å$^2$. The reproducibility of the data at high dose indicates that the data acquired at lower dose are a reliable measure of the state of the sample as deposited.

## 3. Beam Damage Analysis

Figure S3 shows the effect of exposure to electrons on the diffraction signal and 4D STEM orientation mapping results. Figure S3(a) shows the azimuthally averaged parallel beam TEM diffraction intensity as a function of radial k magnitude at accumulated electron doses of 0.24 e$^-$/Å$^2$ and 100.05 e$^-$/Å$^2$. The intensity of the peak at 0.286 Å$^{-1}$ decreases and its width increases with accumulating damage, and its position shifts slightly to the left. The inset figure is a TEM diffraction pattern acquired at an accumulated dose of 0.24 e⁻/Å² from which the azimuthally averaged intensity profile is obtained. Figure S3(b) shows the decay of the normalized signal-to-background ratio of the peak intensity as a function of accumulated electron dose. The measured characteristic dose ($D_{1/e}$) for this molecule is 43.5 e⁻/Å² (0.07 C/cm²), which is typical for such aromatic systems.[14] The dose accumulated in a single 4D STEM nanodiffraction pattern acquired with a 2 nm sized probe with a dwell time of 125 μs is 1.49 e⁻/Å², and the dose accumulated for a 4D STEM scan performed with a 2 nm sized probe and 0.2 nm probe step is 117 e⁻/Å². The collected data from such a scan has an integration of the signal in the dose range of 0-117 e⁻/Å². Figures S3 (c)-(e) show distributions of the fit parameters corresponding to figures 1 (d) and 1 (e).

# 4. Orientation maps and corresponding orientation correlation plots for multiple datasets from each sample

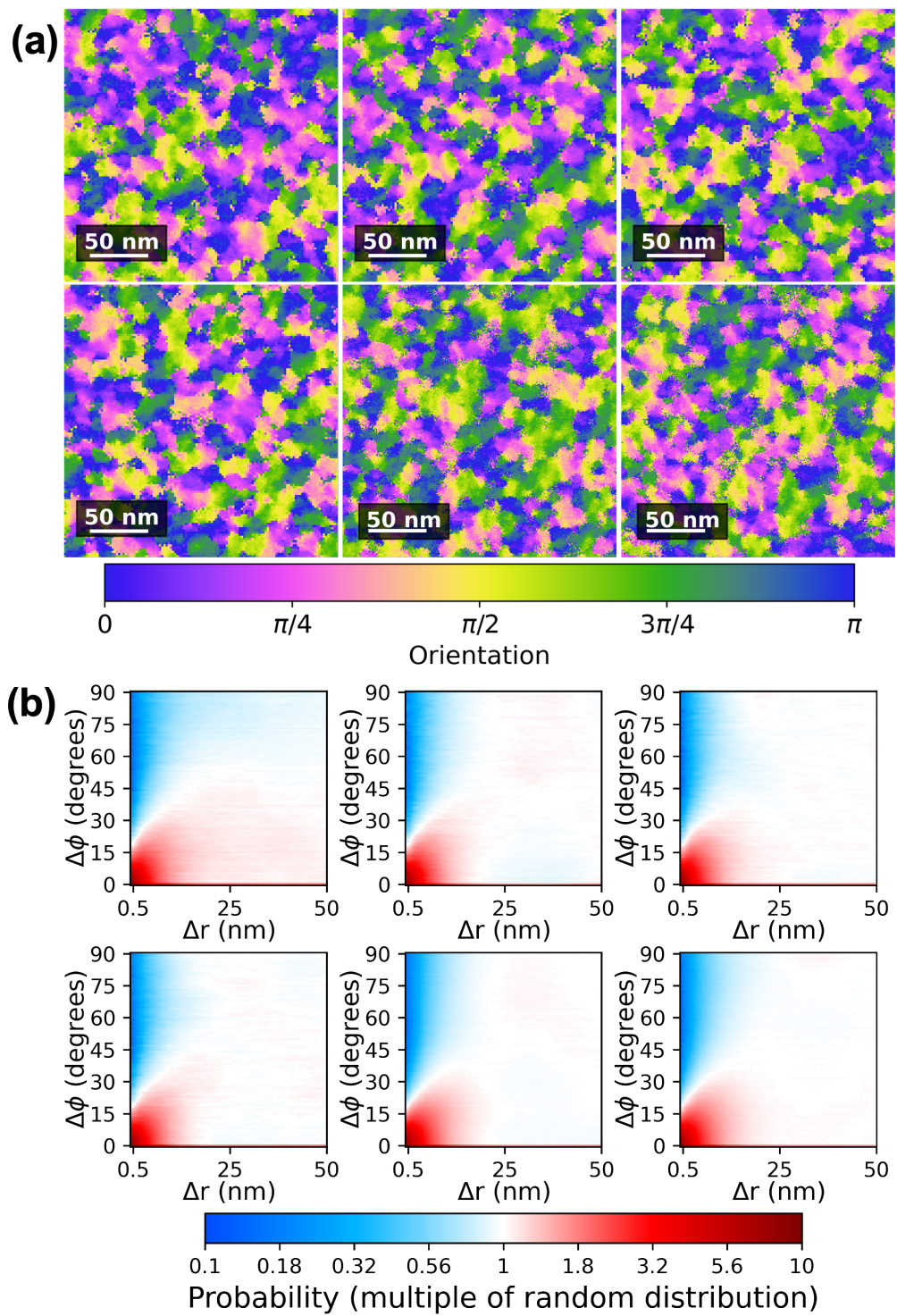

**Supplementary Figure 4:** (a) Orientation maps and (b) corresponding orientation correlation plots for $T_s = 370$ K sample

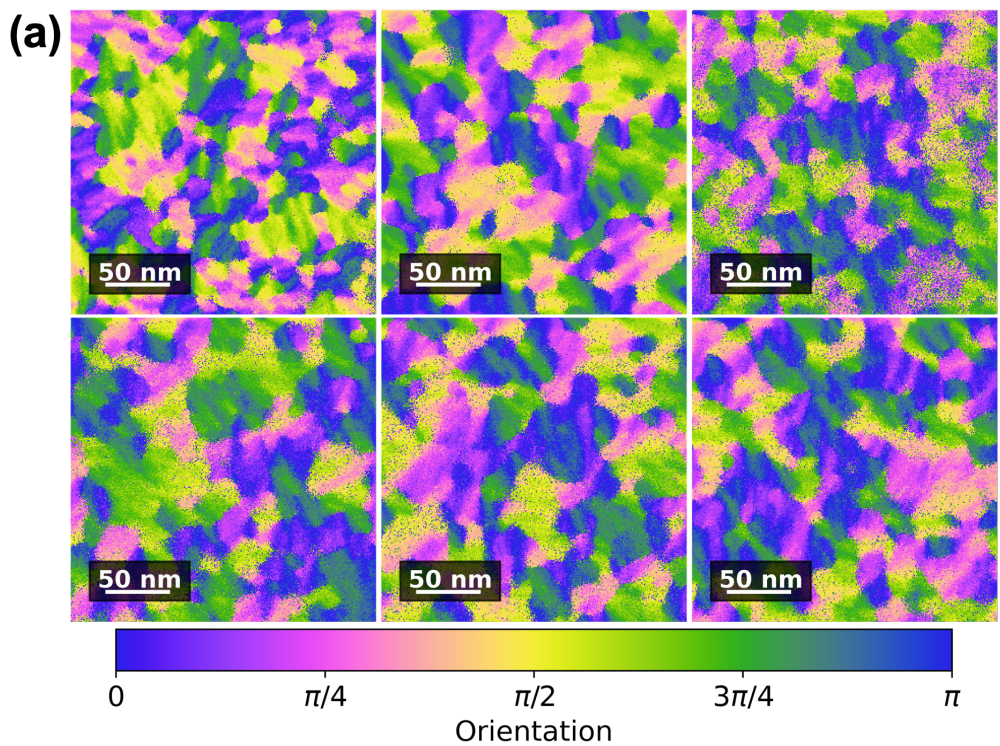

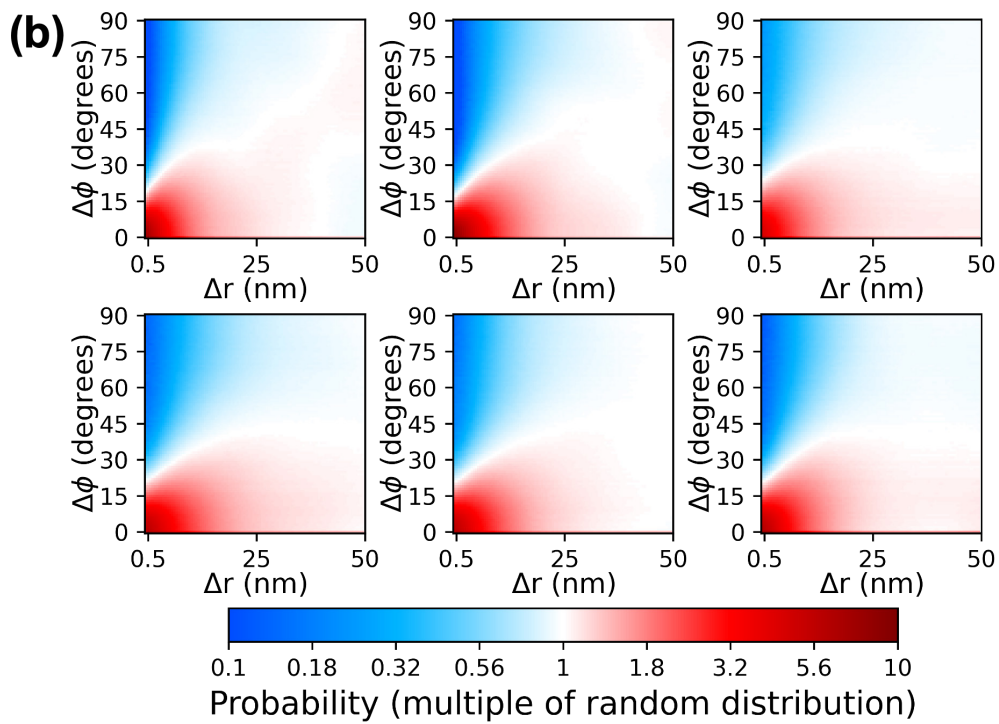

**Supplementary Figure 5:** (a) Orientation maps and (b) corresponding orientation correlation plots for $T_s$ = 380 K sample

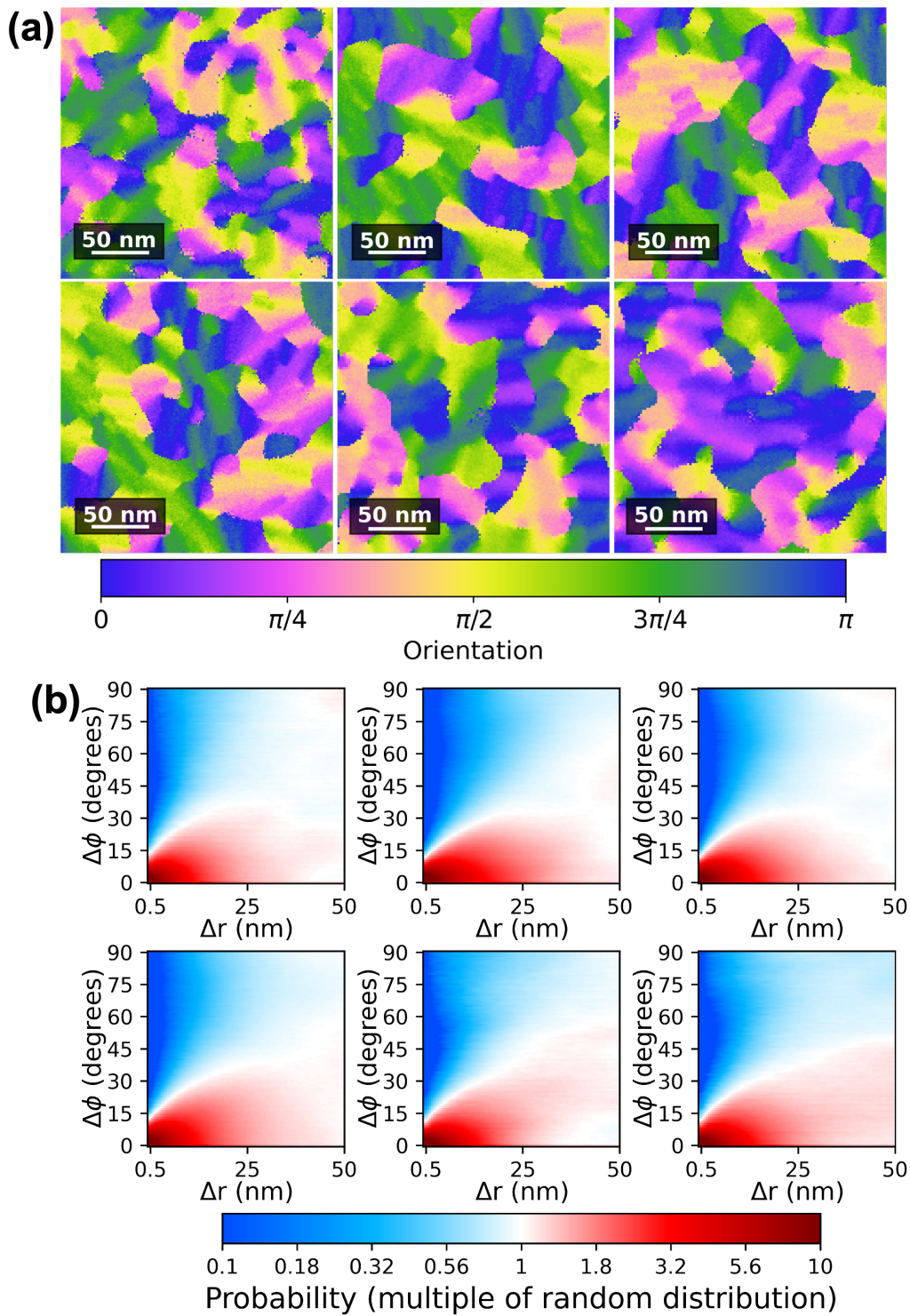

**Supplementary Figure 6:** (a) Orientation maps and (b) corresponding orientation correlation plots for $T_s$ = 392 K sample

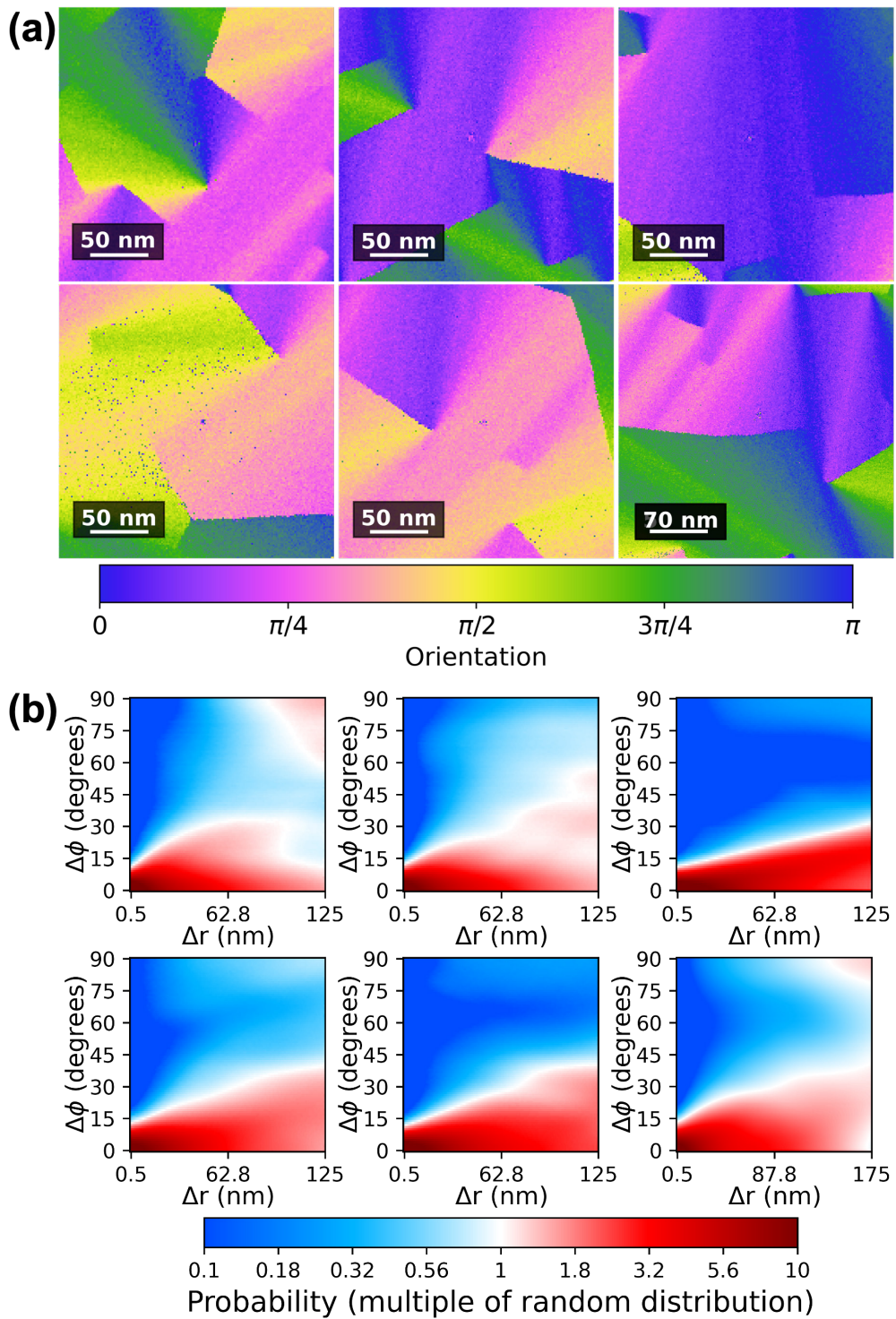

**Supplementary Figure 7:** (a) Orientation maps and (b) corresponding orientation correlation plots for the liquid cooled sample

## 5. TEM diffraction determination of correlation length

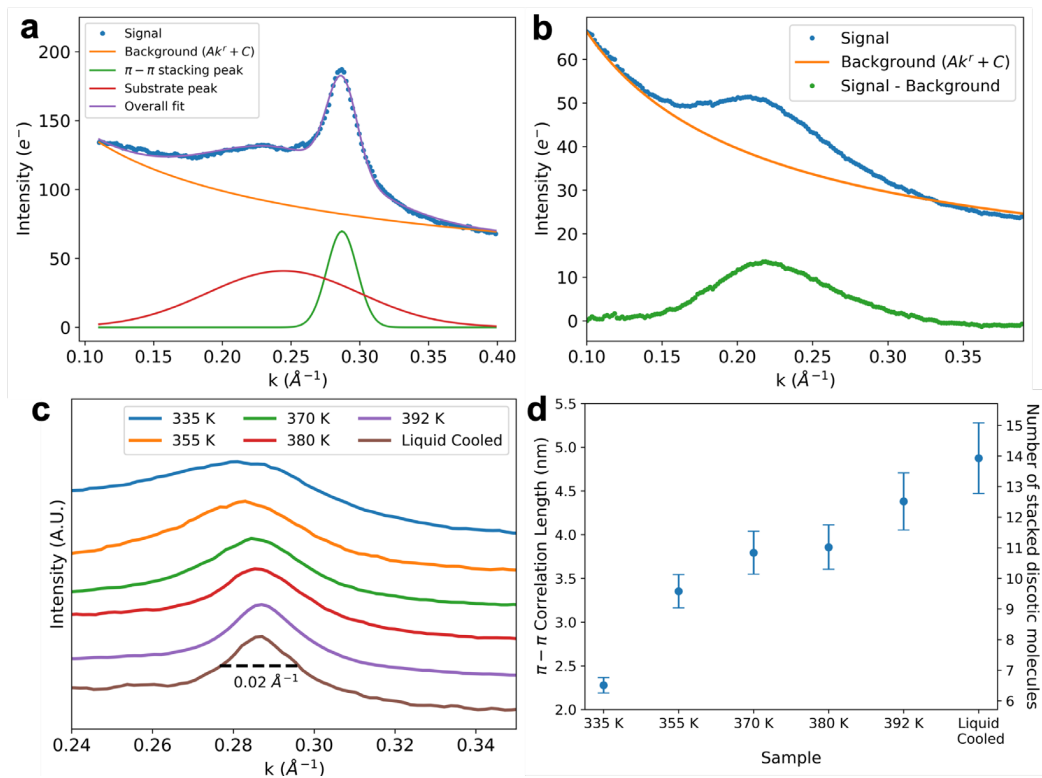

**Supplementary Figure 8:** (a) Illustration of fitting the intensity profile to a power law background and two Gaussians to decompose the signal arising from the substrate and the sample. (b) Intensity profile from a substrate without any film deposited. (c) Background subtracted azimuthal integrated TEM diffraction intensity profiles as a function of radial k for $T_S$ = 335 K, $T_S$ = 355 K, $T_S$ = 370 K, $T_S$ = 380 K, $T_S$ = 392 K, and liquid-cooled glass samples. (d) π–π correlation lengths for all the samples determined from the inverse of the FWHM of the profiles in (c). The right y-axis shows the correlation length in terms of number of stacked discotic molecules.

The inverse width $1/\Delta k$ of a diffraction feature in wide-area diffraction is often used as an ordering length scale for the diffraction feature.[15] For phenanthroperylene ester $1/\Delta k$ of the π–π stacking diffraction feature defines a length over which the π–π distances are well correlated. In nanodiffraction data like in figure 1(c), however, $\Delta k$ is dominated by the probe convergence angle required to create a small probe. The convergence angle is 0.7 mrad, which is 0.056 Å$^{-1}$, and almost exactly the measured $\Delta k$. Figure S8(c) shows profiles through the π–π stacking diffraction feature from parallel beam TEM diffraction with very low convergence on samples vapor deposited at $T_S$ = 335 K, $T_S$ = 355 K, $T_S$ = 370 K, $T_S$ = 380 K, $T_S$ = 392 K, and the liquid-cooled glass sample.

The profiles are background corrected as demonstrated in figure S8(a), but not converted into true structure factors. The π–π stacking correlation lengths are shown in Figure S8(b), along with uncertainties propagated from assuming that the uncertainty in the full width at half maximum is one pixel (0.0017 Å$^{-1}$). The correlation lengths correspond to 6-14 molecules as a function of substrate temperature. They cover the same range as the correlation lengths from x-ray scattering measurements on similar samples, and are nearly in quantitative agreement, especially for the liquid cooled sample.[3] The modest discrepancy (~ 1 nm) may arise from the difference in sample thicknesses and diffraction geometry or uncontrolled factors in the TEM measurements like inelastic scattering.